**Title: SAMRI-2: A Memory-based Model for Cartilage and Meniscus Segmentation in 3D MRIs of the Knee Joint**

**Authors**: Danielle L. Ferreira PhD[1], Bruno A. A. Nunes PhD[1*], Xuzhe Zhang[1,2], Laura Carretero Gomez PhD[3,4], Maggie Fung[5], Ravi Soni PhD[1]

**Affiliations**: [1]GE HealthCare, San Ramon, United States; [2]Columbia University, New York, USA; [3]GE HealthCare, Munich, Germany; [4]LAIMBIO, Rey Juan Carlos University, Madrid, Spain; [5]GE HealthCare, New York, USA.

**Institution**: GE HealthCare, Department of Artificial Intelligence and Machine Learning, 2623 Camino Ramon, San Ramon, CA 94583, USA.

**Abstract**

Accurate morphometric assessment of cartilage—such as thickness/volume—via MRI is essential for monitoring knee osteoarthritis. Segmenting cartilage remains challenging and dependent on extensive expert-annotated datasets, which are heavily subjected to inter-reader variability. Recent advancements in Visual Foundational Models (VFM), especially memory-based approaches, offer opportunities for improving generalizability and robustness. This study introduces a deep learning (DL) method for cartilage and meniscus segmentation from 3D MRIs using interactive, memory-based VFMs. To improve spatial awareness and convergence, we incorporated a Hybrid Shuffling Strategy (HSS) during training and applied a segmentation mask propagation technique to enhance annotation efficiency.

We trained four AI models—a CNN-based 3D-VNet, two automatic transformer-based models (SaMRI2D and SaMRI3D), and a transformer-based promptable memory-based VFM (SAMRI-2)—on 3D knee MRIs from 270 patients using public and internal datasets and evaluated on 57 external cases, including multi-radiologist annotations and different data acquisitions. Model performance was assessed against reference standard using Dice Score (DSC) and Intersection over Union (IoU), with additional morphometric evaluations to further quantify segmentation accuracy.

SAMRI-2 model, trained with HSS, outperformed all other models, achieving an average DSC improvement of 5 points, with a peak improvement of 12 points for tibial cartilage. It also demonstrated the lowest cartilage thickness errors, reducing discrepancies by up to threefold. Notably, SAMRI-2 maintained high performance with as few as three user clicks per volume, reducing annotation effort while ensuring anatomical precision. This memory-based VFM with spatial awareness offers a novel approach for reliable AI-assisted knee MRI segmentation, advancing DL in musculoskeletal imaging.

**Introduction**

Morphometric assessment of cartilage structure, such as thickness measurements, through magnetic resonance imaging (MRI) yields accurate measurements on the progression of Osteoarthritis (OA). These quantitative evaluations are essential for tracking disease onset, monitoring treatment response, and informing clinical decisions. These quantitative measurements depend on effective image processing and segmentation techniques. However, accurate morphometric assessment relies on precise image segmentation, a task that remains challenging due to the complex anatomy and variability in MRI contrast[1]. Recent advances in both MRI acquisition and deep learning (DL)-based techniques have created opportunities for automating such tasks[2–5] and quantitative image analysis has the potential to enhance our understanding of the pathophysiologic changes in cartilage and bone associated with OA.

Compared to automatic segmentation, interactive DL segmentation methods give radiologists more control over the segmentation process[6,7], including in 3D knee MRIs[8]. This control can help minimize inter-observer variability in several key ways. First, standardized user input, such as placing clicks or corrective prompts, reduces subjectivity in boundary definition, ensuring greater consistency across users. Additionally, the AI model can dynamically refine segmentations based on user input, reducing variability that might arise from manual tracing by different radiologists.

The recently introduced Segment Anything Model 2 (SAM2)[9] extends segmentation capabilities to video inputs by introducing a memory mechanism that can retain temporal information from past predictions. Memory-based context utilization further ensures consistency, as contextual information across slices is stored, allowing segmentation decisions to remain stable throughout the volume, even with minor user input changes. This opens an opportunity to apply memory-based Vision Foundational Models (VFMs), such as SAM-2, to the challenging task of segmenting cartilage and meniscus in the knee joint using 3D MRIs, aiming to surpass existing specialist models in performance with superior generalization[10,11].

Inspired by the observation that the video segmentation capability of SAM2 pipeline can be directly applied to 3D image volumes[10], we propose to leverage memory-based VFMs for soft tissue segmentation. To further enhance spatial awareness and learning efficiency, we introduce a Hybrid Shuffling Strategy (HSS), which maintains spatial consistency along the z-axis by shuffling data at the level of sub-volumes—contiguous "chunks" of slices—rather than individual slices. This approach improves segmentation accuracy by reinforcing spatial consistency across slices, allowing the model to better capture anatomical structures and segmentation context. Specifically, we aim to:

(i) Develop accurate segmentation models for morphometric assessment of the femoral, tibial, and patellar cartilages, as well as the four meniscus horns, leveraging memory-based VFMs and enhancing spatial awareness through a Hybrid Shuffling Strategy;

(ii) Compare 3D convolutional neural networks (CNNs) with 2D per-slice VFMs, assessing how incorporating memory mechanisms and structured 3D information—such as chunk-based processing—improves segmentation quality;

(iii) Assess model generalizability by testing performance on independent external datasets acquired using different MR scanners, acquisition protocols, and patient populations, including varying osteoarthritis severity;

(iv) evaluate prompt strategies for assessing 'prompting effort' i.e., the minimum number of human inputs required to achieve high segmentation accuracy while reducing annotation effort.

**Results**

In this study, we compared 3D-VNet, a CNN-based model, with transformer-based approaches: a fully automatic 2D and 3D SAM-based model (without prompts) and the SAM2-based promptable model. The fully automatic SAM-based model relied exclusively on attention mechanisms for segmentation, without human input, providing a transformer-based counterpart to the traditional CNN approach of 3D-VNet, which represents the state-of-the-art (SOTA) in knee MRI segmentation. In contrast, the SAM2-based promptable model introduced interactive segmentation with human-in-the-loop prompting, allowing us to assess its impact on both accuracy and efficiency in clinical applications.

Model evaluation was conducted on two external datasets, D7 and D50. The D50 dataset, a holdout set originally used in the K2S Challenge[12], allowed for a direct comparison with state-of-the-art methods. The D7 dataset, independently manually segmented by two radiologists with over 20 years of experience (RadA and RadB) from two distinct healthcare institutions, was used in our study to provide context for our quantitative results by examining inter-observer variability.

To ensure a fair and systematic comparison, we first evaluated the performance of 3D-VNet as the benchmark. Next, we compared it to both 2D and 3D SAM-based models, isolating the effects of transformer-based architectures in segmentation tasks without human input. By incorporating a 3D decoder into the SAM model, we investigated the potential benefits of capturing spatial features within 3D MRIs. Finally, we introduced the SAM2 model with prompting to examine whether interactive segmentation offers a significant improvement in segmentation performance.

Our results showed that prompting alone did not lead to substantial improvements over the SAM-based model without prompting. Instead, the key contributor to the improved performance of the SAM2 model was the Hybrid Shuffling Strategy (HSS), which enhanced the model's ability to capture spatial dependencies across 3D volumes. The introduction of HSS restructured the volumetric data, improving model convergence and spatial awareness. Without HSS, the memory encoder struggled to effectively handle the complex 3D MRI data, underlining its critical role in optimizing transformer-based segmentation models for knee MRI analysis.

This experimental design enabled us to rigorously assess the impact of model architecture, prompting, and training strategies, highlighting the essential contribution of HSS in enhancing segmentation performance in clinical settings.

*Evaluation of Hybrid Shuffling Strategy (HSS)*:

To assess the impact of the HSS on segmentation performance, we compared the results of SAMRI-2 trained with and without HSS on the D7 dataset. Table 1 presents the Dice Similarity Coefficient (DSC) and Intersection over Union (IoU) for two sets of reference standard annotations in the D7 dataset. The column labeled 'RadA vs RadB' shows inter-observer variability, indicating the agreement between two radiologists (RadA and RadB). The 'Baseline' column reports results for SAMRI-2 trained without Hybrid Shuffling, while the 'Hybrid Shuffle' column highlights the improvements achieved with our Hybrid Shuffling Strategy (HSS). Prompts for SAMRI-2 inference were extracted from the correspondent radiologist's manual segmentations, i.e., if dice score was computed against RadA, click prompts were extracted from the RadA reference standard. In each table, the color scale ranges from dark green (indicating the largest values) to dark red (indicating the smallest values), with shades of yellow and orange representing intermediate values.

The model trained with HSS demonstrated a significant improvement in performance, achieving the highest DSC scores across all tissue types. On average, there was a 5-point improvement in DSC, with a peak improvement of 12 points observed for the tibial cartilage (RadA). Notably, D7 is a particularly challenging dataset, as the inter-radiologist agreement was moderate, with an overall DSC of 0.765. This underscores the effectiveness of HSS in enhancing model performance, even in the presence of significant variability in the reference standard annotations.

*Qualitative Assessment:*

**Figure 2** compares segmentation outputs from the reference standard (RS) and all AI models, highlighting their performance in both axial and sagittal views. It demonstrates each model's ability to capture key knee joint structures, including the femoral, tibial, and patellar cartilages, and the menisci. The figure also allows for an assessment of spatial accuracy and consistency, particularly in the peripheral regions where models exhibit varying performance. Regarding the femoral cartilage, **Figure 2** highlights subtle differences in coverage between the reference standard and AI models, particularly in the shape of the rendered surface. 3D-VNet, SaMRI2D, and SaMRI3D all detect a potential flaw in the femoral cartilage that is not present in the RS rendering. In contrast, SAMRI-2 closely matches the RS rendering, providing more accurate segmentation of the femoral cartilage surface. These findings underscore the variation in model performance, particularly in capturing fine anatomical details.

*Quantitative, and Morphometric Assessment of Cartilage:*

The performance of the models was quantitatively assessed using DSC, IoU and Cartilage thickness measures across different datasets and cartilage compartments. **Table 2** shows that SAMRI3D and SAMRI2D models consistently outperform 3D-Vnet. SAMRI-2 achieves the highest performance across all datasets and compartments, with DSC values ranging from 0.722 to 0.854 and an average of 0.809 for RadA, surpassing other AI models by 5.7 dice points. SAMRI-2's average DSC is 0.831 against RadB, 4.3 points higher than the second-best model. Despite the challenging D7 dataset, where even expert radiologists showed moderate agreement (average DSC of 0.765), SAMRI-2 performed closest to human inter-reader variability, especially for femoral cartilage. SAMRI-2 also outperformed other models on the larger D50 dataset, with a maximum gain of 7 dice points over the second-best model, 3D-Vnet.

In the morphometric assessment of cartilage thickness, also shown in **Table 2**, all models exhibited minimal errors (0.1–0.3mm), with SAMRI-2 showing the closest measurements to reference standard, followed by SaMRI3D. This suggests that incorporating 3D image modulation into VFMs improves segmentation quality, potentially enhancing OA progression measurements. SAMRI-2 achieved the lowest errors across all compartments in both the D7 and D50 datasets, except for the patella in D7. Notably, its average error in D50 was three times lower than the other models.

*Mask Propagation:*

The mask propagation across slices in SAMRI-2 utilizes a strategy where the prediction from one slice is used to generate prompt tokens—comprising both mask predictions and click prompts—for the subsequent slice to be segmented. This process is enhanced by the memory bank mechanism[9], which stores and retrieves relevant features from previous slices to inform and guide the segmentation of the next slice. **Figure 3**, shows how the model performs when prompted on only a few slices, propagating the mask information across the image volume, or when prompted on all slices. The same prompt strategy outlined in Section *Multi-click prompts* was used in all SAMRI-2 inferences. This approach allows SAMRI-2 to maintain segmentation consistency across the entire volume, even with sparse input prompts. SAM2 without fine-tuning performed poorly with prompts on all slices. Our findings align with previous literature, which has reported varying levels of performance for SAM and SAM2 across different medical imaging tasks[13–15]. Specifically, SAM-based models have shown limitations in segmenting small, low-contrast, and poorly delineated structures, all of which characterize cartilage in knee MRIs. Unlike natural images, medical images often lack clear edge information and RGB color, making segmentation more challenging. For example, prior studies reported Dice scores as low as 8.2 for spinal cord segmentation, 13.8 for pleural effusion, and 32.4 for MR glottis[15], reinforcing that SAM's effectiveness varies significantly across different anatomies. Additionally, SAM2's performance remains comparable to that of SAM in several tasks unless fine-tuned for a specific domain[14,16]. While SAM2 extends SAM to promptable video segmentation without sacrificing its image segmentation ability, its default version struggles with complex medical segmentation tasks[14]. The underperformance observed in our study highlights the need for domain-specific adaptations and fine-tuning to improve its applicability to knee cartilage segmentation.

**Discussion**

Interactive segmentation has the potential to revolutionize clinical workflow by enhancing annotation accuracy and fostering greater trust and reliability, while increasing annotation speed and throughput. It provides radiologists with more control over the segmentation annotation process, allowing them to verify the model's output and intervene when necessary, ensuring that the results align with their clinical judgment, while minimizing inter-observer variability - a prevalent issue in manual annotations for medical imaging segmentation[17]. In this study, we propose a 3D knee cartilage and meniscus segmentation method, leveraging an interactive, memory-based visual foundation model (VFM) trained with a Hybrid Shuffling Strategy (HSS). This method improves model convergence and spatial awareness. Our findings suggest that SAMRI-2 generalizes well across different MRI sequences, demonstrating robustness in various clinical contexts.

**Comparison with existing studies:** A number of deep learning-based methods have been proposed for cartilage segmentation in knee MRI, notably in the Knee Cartilage Segmentation (K2S) Challenge[12]. The challenge benchmarked 3D CNN architectures, such as V-Net, nnUNet, and U-Net variations (both 2D and 3D), achieving Dice Similarity Coefficients (DSC) between 0.798 and 0.904 for femoral cartilage, 0.756 and 0.899 for tibial cartilage, and 0.796 and 0.910 for patellar cartilage on the D50 dataset. The best results for cartilage thickness measurements were as follows: femoral: $0.088 \pm 0.07$ mm, tibial: $0.036 \pm 0.09$ mm, and patellar: $0.114 \pm 0.13$ mm. These models were specifically trained and evaluated on a single MRI sequence, optimizing performance in a controlled setting.

In contrast, our study used the D50 dataset exclusively as a holdout test set, without training the model on the same MRI sequence. Despite this, SAMRI-2 achieved performance comparable to the K2S Challenge winners, highlighting its strong generalization capabilities even when applied to datasets with different acquisition protocols.

**Extending the Scope with Transformer-Based Models:** While previous studies have primarily focused on CNN-based approaches, our work extends this by exploring transformer-based models for knee cartilage segmentation. Transformers offer advantages over traditional CNNs, particularly in capturing long-range dependencies and improving spatial awareness. Our study demonstrates the potential of transformer-based segmentation models, particularly when enhanced with HSS to better handle 3D MRI data.

**Minimizing User Interaction for High Accuracy:** To address the challenge of minimizing annotator input while maintaining high segmentation accuracy, we enhanced the SAMRI-2 model with a segmentation propagation method. This approach allows for full-volume segmentation with fewer user interactions, significantly reducing the number of prompts required from radiologists. The effectiveness of this method is partially attributed to the use of a memory bank, which stores features from related slices. This enables the model to leverage contextual information across the volume, improving the efficiency of mask propagation. As a result, our method streamlines the segmentation process, making it more practical for clinical use.

**Limitations and Future Directions:**

A noteworthy limitation of our model is that the iterative click-prompt segmentation was evaluated using a predefined prompt placement strategy derived from reference standard segmentations, which may have influenced the results. Since user input can introduce variability, further assessment through a dedicated reader study would be valuable. This would provide a more realistic evaluation of the model's robustness in clinical settings, where prompts are placed directly by human specialists rather than being inferred from manual annotations. Understanding the model's adaptability to real-world variability will be essential for establishing interactive segmentation as a reliable clinical tool while minimizing the impact of inter-observer variability.

**Materials and Methods**

*Imaging Data:*

The present retrospective study utilized one public dataset (DESS, from Siemens) and one internal dataset (CUBE, from GE Healthcare) for training the AI models. Model evaluation was conducted on two external datasets (D7 and D50). All datasets used in this study were de-identified by the originating hospitals and research institutes. Approval for their use was obtained from the institutional review board, and informed consent was provided by all patients. In the following, we present each dataset and its role in model development. **Table 3** summarizes patient demographics and key dataset characteristics.

*Pre-training – DESS* dataset: The DESS dataset comprises double-echo steady-state (DESS) images from Siemens, from the Osteoarthritis Initiative (OAI) (https://nda.nih.gov/oai), a longitudinal study on osteoarthritis progression. These images were acquired between 2004 and 2006[2,18]. Originally described in[18], and later used in a knee segmentation challenge[2], the dataset includes 176 three-dimensional (3D) MRI volumes, from 88 patients, and was used in our work to pre-train all evaluated models. Patients were scanned using 3-T Magnetom Trio scanners (Siemens Medical Solutions) and quadrature transmit/receive knee coils (USA Instruments) with DESS parameters as follows: field-of-view (FoV) – 140 mm; resolution – 0.36 mm x 0.46 mm x 0.7 mm, zero-filled to 0.36 mm x 0.36 mm x 0.7 mm; echo time – 5 msec; repetition time – 16 ms; and 160 slices.

Manual 3D segmentation masks for four tissue compartments—patellar, tibial, and femoral cartilages, as well as the menisci—were generated by a single expert reader[2]. The DESS dataset has been extensively studied, primarily using this specific sequence.

*Fine-Tuning – CUBE* dataset: Dataset featuring subjects with and without OA, as well as individuals with anterior cruciate ligament (ACL) injuries, as detailed in[19]. A subset of 399 studies from 182 unique patients with manual segmentation masks, selected from a pool of 1435 cases, was used in this study. The remaining cases lack segmentation annotations. Images were acquired using high-resolution 3D fast spin-echo (FSE) fat-suppressed CUBE sequences with the following parameters: FoV - 140 mm (in-plane); repetition time/echo time (TR/TE) – 1500/26.69 ms; matrix – 384 x 384 pixels; image dimensions – $512 \times 512 \times 200$; slice thickness – 0.5mm; bandwidth – 50.0kHz.

All images were acquired with five 3-T MRI scanners (GE Healthcare, Waukesha, WI) and eight surface coils, across different scanners from 2006 to 2018. Over this period, the images were manually annotated by five attending radiologists with over 10 years of experience (YOE) and two trainees from different healthcare institutions[19].

The dataset was originally used by Nunes et al.[19], for anomaly and object detection but was not evaluated for segmentation quality or morphometric measurements, as done in this study.

*External holdout-set – D7* Dataset: Composed by seven clinical cases of high-resolution 3D FSE CUBE acquired between 2018 and 2019 with similar acquisition parameters presented in the *CUBE* dataset description above. Each case was independently segmented by two radiologists (RadA and RadB), each with over 20 YOE, from separate healthcare institutions. No pathology readings were available for this dataset. Figure 4 illustrates the agreement between the two radiologists, and shows sagittal slices for two cases overlayed by their correspondent segmentations from RadA and RadB. Case 2 was the case for which radiologists presented the worst agreement, while Case 6 presented the best agreement amongst the 7 cases in the dataset.

In our study, the D7 dataset was used to assess inter-observer variability and provide context for our quantitative results. To our knowledge, it has not been reported in previous studies.

*External holdout-set – D50* Dataset: The D50 dataset consists of 50 high-resolution 3D FSE CUBE sequences acquired with parameters similar to those of the CUBE dataset used for training. These cases were manually annotated by a radiologist with over 20 YOE, and no pathology readings were available. Scans were performed on a GE Discovery MR750 3T Scanner (GE Healthcare, Milwaukee, WI) using an 18-channel knee transmit/receive coil and the following acquisition parameters: FoV - 150mm (in-plane); repetition time/echo time (TR/TE) – 1002/29 ms; matrix – 256 x 256 x 200 pixels; image dimensions – $512 \times 512 \times 200$; slice thickness – 0.6mm; bandwidth – 244.0kHz. This dataset includes annotations for femoral, tibial, and patellar cartilage, but no manual meniscus segmentations were provided[12].

Originally introduced in the K2S Challenge[12], the D50 dataset was used as a test set containing only cartilage segmentations. The K2S research group carefully selected these 50 cases from a larger pool of 816. The challenge focused on a specific MRI sequence, and models were trained and evaluated accordingly. In our study, we used the D50 dataset exclusively as a holdout set. Despite this, our model achieved performance comparable to the K2S Challenge winners, even without training on the same MRI sequence, demonstrating its ability to generalize to an unseen dataset.

*Deep Learning Models for Knee MRI Segmentation:*

Four deep learning models were trained to learn segmentations of soft tissue structures in 3D knee MRIs, namely the *femoral*, *tibial*, and *patellar* cartilages, as well as the four *meniscus horns*. Of these, three models were designed as specialist models, while one was trained as a promptable model, potentially aiding in semi-automatic labeling of 3D knee MRIs.

- 3D-Vnet[20] (CNN-based): Extensively used in the Osteoarthritis Imaging Knee MRI Segmentation Challenge (1), where it outperformed other CNN-based architectures for knee MRI segmentation. Due to its strong performance in this challenge, 3D-VNet was selected as a representative CNN-based model for comparison with three recent transformer-based VFMs: two SAM-based methods[21], SaMRI2D and SaMRI3D, and one SAM2-based approach[9], SAMRI-2.

- SaMRI2D (VFM SAM-based): Excludes the prompt decoder, enabling fully automated segmentation without human input. This design aims to compare the performance of SAM-based transformer architectures (which use attention mechanisms) in knee MRI segmentation tasks without prompts, against 3D-VNet, which is also trained to perform fully automated segmentation. The goal is to isolate the differences in performance that arise from the architectural choices (CNN vs. transformer) rather than the use of prompts.

- SaMRI3D (VFM SAM-based): Modifies the SAM architecture, originally designed for 2D segmentation, to incorporate 3D spatial information using cross-view attention (2) in the mask decoder. This modification enables SaMRI3D to handle 3D volume data. The prompt encoder is excluded to focus solely on the impact of 3D integration, allowing for a direct comparison with 3D-VNet (CNN-based) and its 2D version, SaMRI2D. This design enables us to assess whether the attention mechanism in SaMRI3D offers any advantages over the more traditional CNN-based approach in 3D-VNet and the 2D segmentation method in SaMRI2D.

- SAMRI-2 (VFM SAM-based): Trained end-to-end, incorporating the image, memory, and prompt encoders, along with the mask decoder. We fine-tuned it using the sam2_hiera_tiny weights from the smaller 'tiny' SAM2 architecture. However, training the memory encoder alone was insufficient for convergence on 3D spatial data. Convergence improved significantly with the introduction of a **Hybrid Shuffling Strategy (HSS)**, discussed in the next section.

For the promptable model, we employed a Multi-click point prompt strategy, detailed below.

*Hybrid Shuffling Strategy:*

In SAMRI-2, we adapted the memory mechanism from SAM2 to preserve spatial information along the z-axis by introducing a Hybrid Shuffling Strategy (HSS). The strategy involves sampling sub-volumes, or "chunks", consisting of consecutive slices along the z-axis from the MRI volume. These chunks are fed to the model in batches, and at the end of each epoch, the dataset is shuffled by chunk, rather than by individual slices, as shown in **Figure 1**. The HSS plays a key role in improving the effectiveness of the cross-attention mechanism in the memory attention layer. By ensuring that memory embeddings appended to the memory bank retain spatial correlations between consecutive slices, the HSS helps the model better capture and leverage spatial dependencies. This approach enhances the model's ability to learn coherent spatial patterns while still benefiting from the variability introduced by shuffling.

*Multi-click prompts:*

Point click prompts were used instead of bounding boxes for segmenting knee soft tissue, as bounding boxes are less effective for non-convex shapes and may inadvertently include unwanted anatomical structures, leading to incorrect segmentation. While previous research[22] suggests that box-based prompts can yield relatively higher segmentation accuracy, large bounding boxes may encompass multiple instances and background structures, potentially confusing the model and resulting in incorrect segmentation. Other work[23] supports the use of iterative click prompts for achieving similar results to bounding boxes. We implemented an iterative click prompt technique to simulate realistic user-generated interactions. Initially, the first point is selected from an area near the centroid of the reference standard (RS) mask. Specifically, the (x, y) coordinates are chosen from within the 30th percentile of the largest Euclidean distances to the mask boundary. Subsequent points are determined by calculating an error map between the model's prediction and the RS mask. Points are then selected from regions with the highest error. Each point is labeled as either a positive or negative prompt based on its location relative to the RS mask: a point is classified as a false negative if it falls inside the mask (indicating an error region), and as a false positive if it falls outside the mask. During training, our model was prompted up to eight times per slice, per compartment, with a single click per iteration. During inference, only one click was used, selected at the centroid of the RS mask.

*Training Strategy and Architectures:*

All four DL models were trained with the same data split with the objective of segmenting cartilage and meniscus on 3D MRIs of the knee. The training datasets were randomly split per patient into 80% for training and 20% for validation.

*Pre-processing*: All four models were trained on 3D volumes standardized with the same FoV of 144mm in-plane and 80mm out-of-plane. Training images were augmented using flipping on the z-axis (50% of the time), rotating on all 3 axis on random angles chosen from 15 degrees on xy-axis and 9 degrees on xz and yz axis (70% of the time), random Gaussian noise (50% of the time), random bias field (20% of the time) and random elastic deformation (50% of the time).

*Pre-training*: Models were first pre-trained on the DESS dataset with the objective of learning the geometric features necessary for segmenting the regions-of-interest (ROIs) from quality manual segmentations.

*Fine-tuning*: A second dataset i.e., CUBE, consisting solely of the targeted imaging data, was used for further training. This allowed the models to learn representations across various aspects such as patient anatomy, imaging contrast, and the positions of anatomical structures within the volume.

*Model selection*: Training was stopped on both Pre-training and Fine-tuning when the dice score (DSC) metric for the validation set did not improve for 10 epochs. Final model chosen was the one that presented the highest DSC overall classes.

*Vnet*: The 3D CNN based architecture Vnet[20] was trained on images volumes resized to 288 x 288 x 160 pixels. The model's weights were initialized with the Glorot uniform initializer, and the training was performed with a batch size of 4 due to GPU memory constraints. The loss function combined weighted cross-entropy loss with Dice loss. The Adam optimizer was employed, dropout was not used, and the initial learning rate (lr) was set to $10^{-4}$, with a decay-on-plateau strategy that reduced the lr by half every 5 epochs, down to a minimum of $10^{-6}$ if the validation DSC overall classes did not improve.

*SaMRI2D*: The SAM (3) model was trained end-to-end, with both the image encoder and mask decoder fine-tuned on knee MRI data using the original *sam_vit_base* weights (90M parameters), on 2D slice-by-slice approach. During training, the dataloader loaded MRI volumes, applied augmentation transformations, and randomly selected a slice to resize to (1024 x 1024) pixels before feeding it to the network. A batch size of 2 was used, along with the same learning rate scheduler as employed for 3D-Vnet. The prompt decoder was excluded from the training process, as the goal was to develop a specialized model that operates without human intervention.

*SaMRI3D*: The SAM architecture was modified here to be able to modulate 3D information via cross-view attention (2) on the mask decoder. During training, the model was fed random 3D patches of size (*batch*, h, w, z) = (1, 96, 96, 96)). SaMRI3D was fine-tuned end-to-end using the original pre-trained *sam_vit_base* weights, with all other training parameters and procedures consistent with those used for the previous architectures. As before, the prompt encoder was not utilized in this setup.

*SAMRI-2*: The interactive SAM2 model[9] was trained end-to-end, with the image encoder, prompt encoder, and mask decoder all fine-tuned from the *sam2_hiera_tiny* weights from the smaller 'tiny' architecture. This choice was made to accelerate the training process. During training, the model was fed batches of 8 consecutive slices of size 1024 x 1024. Feeding consecutive slices to the model however was not enough to achieve convergence, which we improved by introducing a *hybrid shuffling* strategy (HSS) during training.

*Model Sizes*: **Table 4** compares the number of parameters across all the architectures and inference time measured while inferring on all 50 cases of D50 dataset. The number of parameters is linked to the model's complexity and inference time. In this context, inference time refers to the duration it takes for the model to output the segmented volume. While the CNN-based model has the lighter footprint and fastest inference time, SAMRI-2 presents a reasonable compromise between model size, inference time and effort needed to achieve quality performance.

*Quantitative Metrics and Statistical Analysis:*

The *Dice Score* (DSC) and *Intersection over Union* (IoU) are widely used for evaluation in medical image segmentation tasks. Both metrics, range from 0 smallest value indicating the worst performance, to the highest value of 1.0, which indicates perfect similarity. DSC is a measure of overlap between the predicted segmentation P and the reference standard segmentation R. It is defined as:

$$DSC = \frac{2|\boldsymbol{P}| \cap |\boldsymbol{R}|}{|\boldsymbol{P}| + |\boldsymbol{R}|}$$

where |P ∩ R| represents the intersection of the predicted and RS segmentation, and |P| and |R| are the cardinalities of the predicted and reference standard sets, respectively.

The IoU on the other hand is defined as:

$$IoU = \frac{|\boldsymbol{P} \cap \boldsymbol{R}|}{|\boldsymbol{P} \cup \boldsymbol{R}|}$$

where |P ∪ R| is the union of the predicted and RS segmentations, representing the total area covered by either of the sets. IoU is a stricter metric than DSC because it penalizes false positives more heavily. It evaluates how much overlap exists between the predicted segmentation and the RS in relation to the total combined area, making it a robust indicator of segmentation quality in various medical imaging scenarios. The absolute error in cartilage thickness measurements was computed following the method described in[24], by comparing the predicted values with those extracted from the reference standard (RS) segmentations.

Specifically, cartilage thickness $T_{ba}$ at a given point $P_b$ on the bone surface is defined as the Euclidean distance to its nearest neighbor $P_a$ on the articular surface:

$$T_{ba} = \text{Distance}(P_b, P_a)$$

where $P_b$ is the reference point on the bone surface, and $P_a$ is its nearest neighbor on the articular surface. Lower absolute errors indicate higher accuracy and closer agreement with the RS.

Statistical significance of differences between models was assessed using the *Wilcoxon rank-sum test* implemented in Python using the *SciPy* package. P-values less than 0.05 suggest significant differences in distribution. Lower errors indicate greater accuracy in the predictions.

**Data Availability**

The minimum dataset is available in the Osteoarthritis Initiative (OAI) repository, https://nda.nih.gov/oai

**Author Contributions**

D.F., B.N., and RS conceived the study. D.F. and B.N. wrote the manuscript and performed data analyses. D.F., B.N., and X.Z. wrote Python code and checked algorithms. L.G. and M.F. performed data acquisition. All authors reviewed the manuscript.

**Additional Information**

**Competing Interests**

All authors were employed by GE Healthcare during the research and manuscript preparation. However, the research was conducted independently, and the authors declare that the results and conclusions are presented objectively, without external influence from their employer.

**Tables:**

**Table 1:** DSC and IoU for the two sets of reference standard in D7. Column 'RadA vs RadB' shows inter-observer variability. Column 'Baseline' shows results for memory-based SAMRI-2 trained without hybrid shuffling, while column 'Hybrid Shuffle' shows the benefit of our strategy. In each one of the four tables, the color scale goes from dark green (largest values), passing by shades of yellow and orange, down to dark red (smallest values).

| D7 | RoI | | Hybrid shuffle | | Baseline | | RadA vs RadB | | | Hybrid shuffle | | Baseline | | RadA vs RadB | |
|---|---|---|---|---|---|---|---|---|---|---|---|---|---|---|---|
| | | | Avg. | std | Avg | Std | Avg | Std | | Avg. | std | Avg | Std | Avg | Std |
| RadA | femur | DSC | **0.722** | 0.075 | 0.67 | 0.068 | 0.795 | 0.05 | IOU | **0.584** | 0.090 | 0.507 | 0.078 | 0.663 | 0.05 |
| | tibia | | **0.822** | 0.038 | 0.695 | 0.051 | 0.768 | 0.08 | | **0.686** | 0.052 | 0.534 | 0.058 | 0.628 | 0.08 |
| | patella | | **0.803** | 0.041 | 0.779 | 0.052 | 0.777 | 0.07 | | **0.696** | 0.058 | 0.641 | 0.068 | 0.639 | 0.07 |
| | meniscus | | **0.854** | 0.068 | 0.832 | 0.015 | 0.719 | 0.06 | | **0.697** | 0.093 | 0.712 | 0.022 | 0.565 | 0.06 |
| | All | | **0.809** | 0.056 | 0.744 | 0.046 | 0.765 | 0.06 | | **0.666** | 0.073 | 0.598 | 0.056 | 0.624 | 0.06 |
| RadB | femur | DSC | **0.779** | 0.061 | 0.745 | 0.037 | 0.795 | 0.05 | IOU | **0.64** | 0.08 | 0.595 | 0.048 | 0.663 | 0.05 |
| | tibia | | **0.862** | 0.025 | 0.805 | 0.022 | 0.768 | 0.08 | | **0.76** | 0.037 | 0.674 | 0.031 | 0.628 | 0.08 |
| | patella | | **0.845** | 0.034 | 0.808 | 0.064 | 0.777 | 0.07 | | **0.73** | 0.051 | 0.682 | 0.087 | 0.639 | 0.07 |
| | meniscus | | **0.837** | 0.023 | 0.812 | 0.026 | 0.719 | 0.06 | | **0.72** | 0.035 | 0.684 | 0.037 | 0.565 | 0.06 |
| | All | | **0.831** | 0.036 | 0.792 | 0.037 | 0.765 | 0.06 | | **0.71** | 0.051 | 0.659 | 0.051 | 0.624 | 0.06 |

**Table 2: DSC, IoU, Cartilage thickness measurements and Absolute error** (AAE) comparing AI models-driven to manual segmentations for Datasets D7 and D50. D7 includes two independent reference standard sets annotated by RadA and RadB, with the column 'RadA vs. RadB' showing interobserver variability (DSC, IoU, and thickness metrics). **Bold numbers** indicate lower error values. Avg = average, Std = standard deviation, DSC = Dice Score, IoU = Intersection over Union. AAE= Average Absolute Error. An asterisk (*) indicates errors larger than inter-rater variability. Statistical significance was assessed using the Wilcoxon rank-sum test: † $p < 0.05$; ‡ $p < 10^{-7}$

| Dataset - D7 | | Human (GT) | Vnet | | SAMRI3D | | SAMRI2D | | SAMRI-2 | |
|---|---|---|---|---|---|---|---|---|---|---|
| | | Rad A vs Rad B | Rad A | Rad B | Rad A | Rad B | Rad A | Rad B | Rad A | Rad B |
| femur | DSC Avg (std) | 0.795 (0.05) | 0.710 (0.08) | 0.767 (0.05) | **0.722 (0.07)** | 0.770 (0.03) | 0.689 (0.084) | 0.754 (0.05) | **0.722 (0.075)** | **0.779 (0.061)** |
| tibia | | 0.768 (0.08) | 0.747 (0.06) | 0.817 (0.03) | 0.746 (0.07) | 0.813 (0.02) | 0.758 (0.063) | 0.827 (0.02) | **0.822 (0.038)** | **0.862 (0.025)** |
| patella | | 0.777 (0.07) | 0.758 (0.09) | 0.788 (0.04) | 0.774 (0.08) | 0.818 (0.04) | 0.753 (0.090) | 0.797 (0.03) | **0.803 (0.041)** | **0.845 (0.034)** |
| meniscus | | 0.719 (0.06) | 0.777 (0.04) | 0.758 (0.05) | 0.772 (0.03) | 0.753 (0.04) | 0.781 (0.037) | 0.759 (0.06) | **0.854 (0.068)** | **0.837 (0.023)** |
| All | | 0.765 (0.06) | 0.748 (0.08) | 0.783 (0.04) | 0.753 (0.06) | 0.788 (0.03) | 0.745 (0.07) | 0.784 (0.04) | **0.809 (0.056)** | **0.831 (0.036)** |
| femur | IoU Avg (std) | 0.663 (0.07) | 0.536 (0.10) | 0.624 (0.06) | 0.569 (0.09) | 0.627 (0.04) | 0.531 (0.10) | 0.608 (0.06) | **0.584 (0.090)** | **0.641 (0.080)** |
| tibia | | 0.628 (0.09) | 0.589 (0.09) | 0.692 (0.04) | 0.598 (0.09) | 0.685 (0.02) | 0.613 (0.08) | 0.705 (0.03) | **0.686 (0.052)** | **0.758 (0.037)** |
| patella | | 0.639 (0.09) | 0.601 (0.12) | 0.653 (0.06) | 0.637 (0.10) | 0.694 (0.06) | 0.611 (0.10) | 0.664 (0.05) | **0.696 (0.058)** | **0.733 (0.051)** |
| meniscus | | 0.565 (0.08) | 0.610 (0.05) | 0.613 (0.07) | 0.629 (0.04) | 0.605 (0.06) | 0.642 (0.05) | 0.615 (0.08) | **0.697 (0.093)** | **0.72 (0.035)** |
| All | | 0.624 (0.08) | 0.584 (0.09) | 0.645 (0.05) | 0.608 (0.08) | 0.653 (0.05) | 0.599 (0.08) | 0.648 (0.05) | **0.666 (0.073)** | **0.713 (0.051)** |
| femur | Thickness AAE (mm) Avg (std) | 0.295 (0.045) | 0.382* (0.206) | 0.182 (0.099) | 0.276 (0.164) | 0.119 (0.057) | 0.347* (0.215) | 0.131 (0.088) | **0.254 (0.023)** | **0.099 (0.035)** |
| tibia | | 0.354 (0.307) | 0.320 (0.260) | 0.129 (0.068) | 0.249 (0.222) | 0.219 (0.14) | 0.300 (0.250) | **0.107 (0.066)** | **0.199 (0.067)** | 0.228 (0.101) |
| patella | | 0.255 (0.124) | 0.332* (0.118) | 0.208 (0.188) | **0.189 (0.147)** | 0.201 (0.084) | 0.265* (0.197) | 0.272* (0.129) | 0.341* (0.050) | **0.178 (0.069)** |
| All | | 0.301 (0.353) | 0.345* (0.195) | 0.173 (0.118) | **0.238 (0.178)** | 0.180 (0.094) | 0.304* (0.221) | 0.170 (0.094) | 0.265 (0.047) | **0.168 (0.068)** |
| femur | Thickness measure (mm) Avg (std) | 1.472 (0.335) 1.751 (0.315) | 1.725 (0.478) | | 1.590 (0.461) | | 1.690 (0.500) | | 1.726 (0.238) | |
| tibia | | 1.576 (0.229) 1.998 (0.299) | 1.821 (0.463) | | 1.754 (0.424) | | 1.822 (0.474) | | 1.775 (0.261) | |
| patella | | 2.134 (0.424) 2.382 (0.459) | 2.288 (0.563) | | 2.145 (0.506) | | 2.221 (0.545) | | 2.475 (0.406) | |
| All | | 1.727 (0.329) 2.044 (0.358) | 2.032 (0.501) | | 1.901 (0.464) | | 2.00 (0.506) | | 1.992 (0.301) | |
| Dataset - D50 | | GT | Vnet | SAMRI3D | SAMRI2D | SAMRI-2 | Vnet | SAMRI3D | SAMRI2D | SAMRI-2 |
| | | DICE - Avg (std) | | | | | IOU - Avg (std) | | | |
| femur | DSC IoU | — | 0.779 (0.03) | 0.709 (0.044) | 0.758 (0.037) | **0.849 (0.024)** | 0.638 (0.037) | 0.612 (0.047) | 0.551 (0.052) | **0.743 (0.036)** |
| tibia | | — | 0.786 (0.035) | 0.682 (0.052) | 0.709 (0.047) | **0.839 (0.033)** | 0.648 (0.047) | 0.551 (0.057) | 0.519 (0.061) | **0.720 (0.047)** |
| patella | | — | 0.847 (0.054) | 0.758 (0.083) | 0.812 (0.071) | **0.858 (0.048)** | 0.738 (0.073) | 0.688 (0.087) | 0.617 (0.095) | **0.756 (0.065)** |
| meniscus | | — | **0.862 (0.012)** | 0.800 (0.039) | 0.804 (0.045) | 0.827 (0.026) | 0.727 (0.023) | 0.674 (0.087) | 0.669 (0.054) | **0.703 (0.037)** |
| All | | — | 0.818 (0.032) | 0.737 (0.055) | 0.771 (0.050) | **0.843 (0.033)** | 0.688 (0.045) | 0.631 (0.063) | 0.589 (0.065) | **0.731 (0.046)** |
| | | | Measurements | | | | AAE - Average Absolute Error | | | |
| femur | Thickness (mm) Avg (std) | 2.115 (0.210) | 1.816‡ (0.152) | 1.806‡ (0.297) | 1.907‡ (0.169) | 2.108† (0.221) | 0.300 (0.144) | 0.308 (0.177) | 0.239 (0.158) | **0.077 (0.053)** |
| tibia | | 2.124 (0.240) | 1.803‡ (0.206) | 1.787‡ (0.382) | 1.877‡ (0.190) | 2.086† (0.212) | 0.321 (0.047) | 0.333 (0.089) | 0.284 (0.164) | **0.106 (0.78)** |
| patella | | 2.733 (0.354) | 2.397‡ (0.354) | 2.473‡ (0.339) | 2.396‡ (0.251) | 2.786† (0.361) | 0.359 (0.085) | 0.277 (0.079) | 0.361 (0.182) | **0.138 (0.093)** |
| All | | 2.324 (0.268) | 2.005‡ (0.237) | 2.022‡ (0.340) | 2.060‡ (0.203) | 2.327† (0.265) | 0.326 (0.092) | 0.306 (0.115) | 0.295 (0.168) | **0.107 (0.075)** |

**Table 3:** Description of all datasets used for training and performance evaluation.

| Datasets | Demographic Distribution mean (stdv) [range] | Nb. Patients | Sample Size | Used For |
|---|---|---|---|---|
| DESS | Age: 58 (9) [45,78]<br>BMI(Kg/m*m): 32 (5)<br>Sex: 38 (51%) out of 74 are Female | 88 | 176 | Pre-training |
| CUBE | Age: 43 (16) [15,68]<br>BMI: 23.28 (5.35) [19,29]<br>Sex: 89 (49%) out of 182 are Female | 182 | 399 | Fine-tuning |
| D7 | Age: 50.6 (14.73) [21,75]<br>Weight(Kg): 83.10 (16.76) [50,99]<br>Sex: 4 (57%) out of 7 are Female | 7 | 7 | Holdout test |
| D50 | Age: 44.54 (14.66) [20,75]<br>Weight: 70.49 (16.78) [44.45, 122.47]<br>Sex: 26 (52%) out of 50 are Female | 50 | 50 | Holdout test |

**Table 4:** Models' comparison – Nb of parameters and Average Inference time in seconds per volume. NvidiaV100 (32GB) GPU. Nb of workers: 1.

| Model | Nb of parameters | Inference Time |
|---|---|---|
| SaMRI3D | 91M | 12.57(0.21) |
| SaMRI2D | 94M | 22.88(0.20) |
| Vnet | 11M | 0.46(0.005) |
| SAMRI-2 | 38M | 4.24(0.397) |

**Figures:**

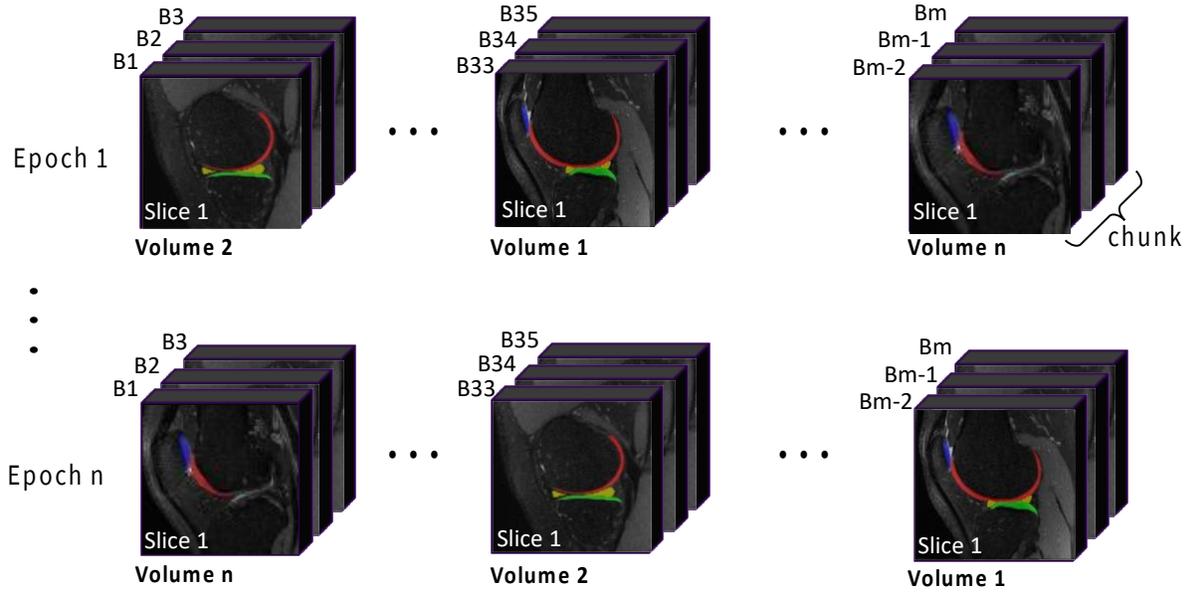

**Figure 1:** Hybrid Shuffling Strategy (HSS) – During training, at the start of each epoch, the model is fed batches of 'S' sequential slices, sampled in order from the same volume. Specifically, in Epoch 1, Batch 1 (B1) includes slices [1, S], Batch 2 (B2) includes slices [S+1, S+S], and so on, until the end of the volume. At the end of each epoch, the chunks are shuffled, and the process repeats in the next epoch.

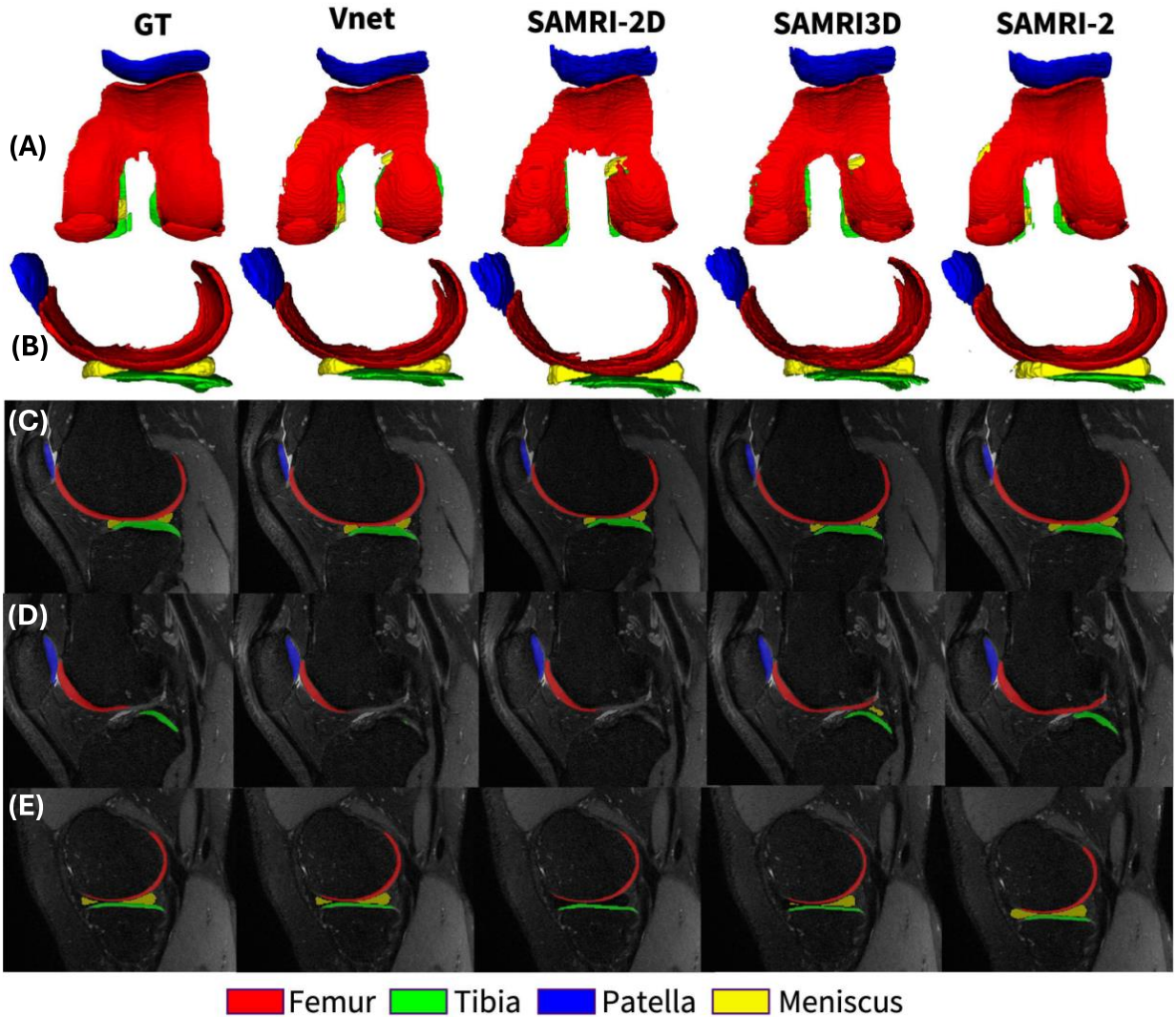

**Figure 2:** 3D rendering of segmentation outputs from the reference standard and all models in axial **(A)** and sagittal **(B)** views. **(C)**, **(D)**, and **(E)** show sagittal slices form the same case with the respective segmentations overlayed. In **(C)**, strong agreement is observed on sagittal slices through the center of the structures. However, in peripheral slices **(D)** and **(E)**, models begin to disagree, particularly for tibial cartilage and the meniscus, though these discrepancies are limited only to a few slices.

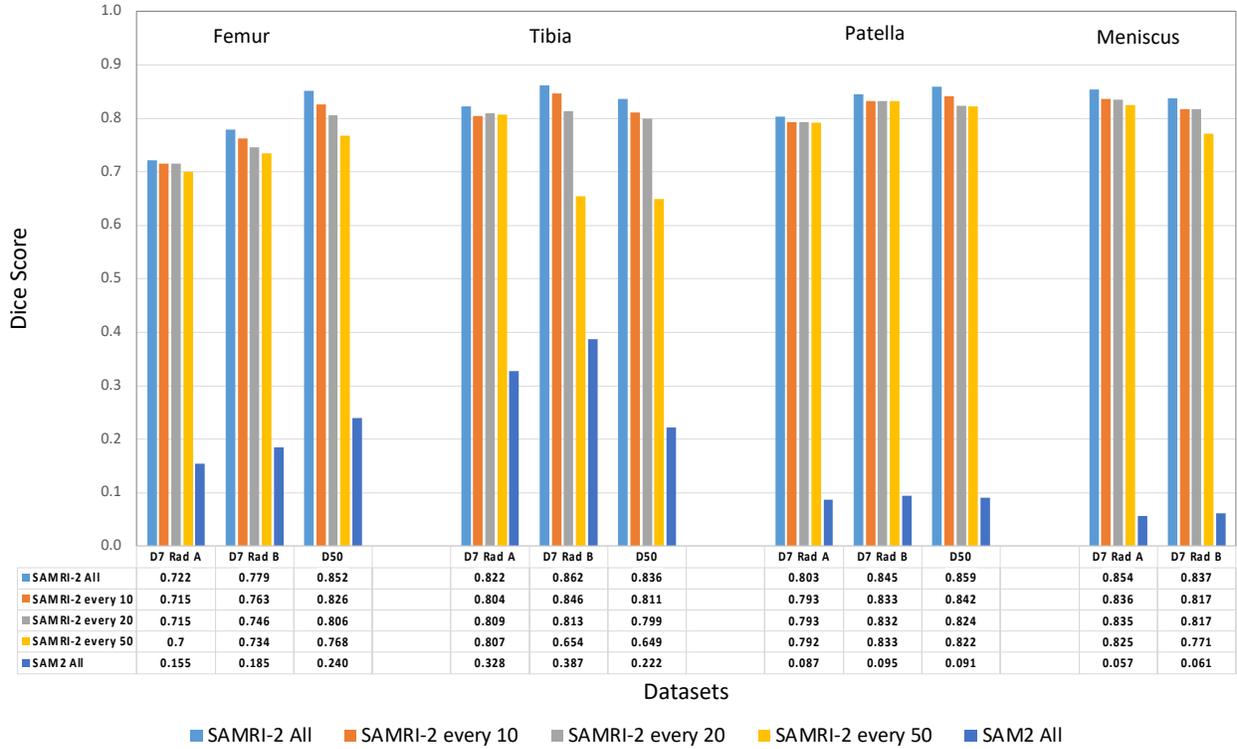

**Figure 3:** Evaluation of mask propagation for SAMRI-2 and SAM2 (without fine-tuning) using different slice selection strategies: (a) placing click prompts on every sagittal slice of the volume (SAMRI-2 ALL), and (b) placing a click prompt on the first slice where the structure appears, then propagating it across the next 10 (SAMRI-2 every 10), 20 (SAMRI-2 every 20), or 50 (SAMRI-2 every 50) subsequent sagittal slices. Four segmentation classes (femoral, tibial, and patellar cartilages, and the meniscus) were evaluated using a single click prompt for each class.

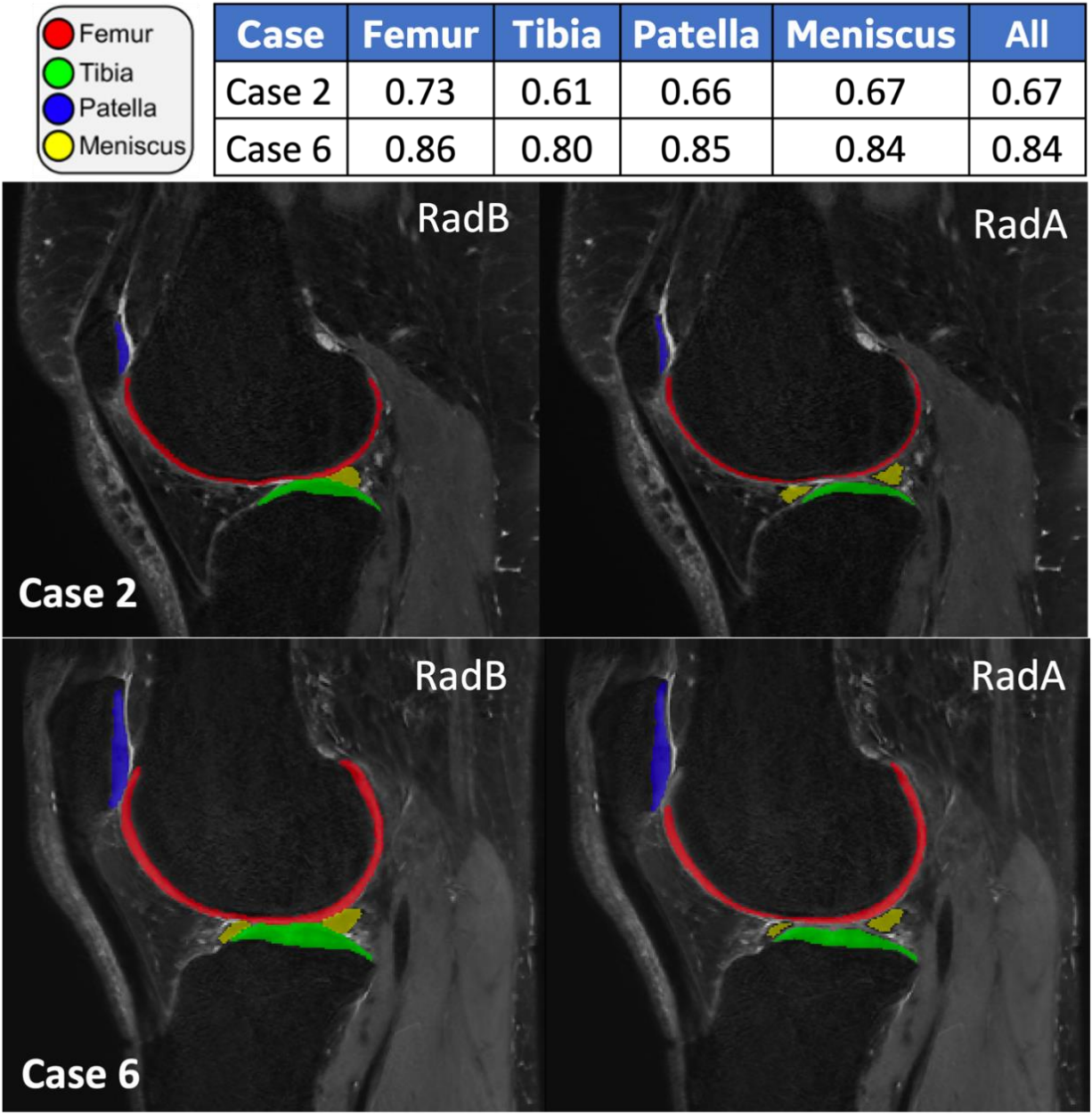

**Figure 4:** Sagittal slices for two cases with the best (Case 6) and the worst agreement (Case 2) between the two radiologists Rad A and Rad B, in the D7 holdout testset. Table shows 3D dice score (computed for the whole volume) for each compartment.

29